\documentclass[12pt, A4]{article}
\usepackage{amsmath}
\usepackage{graphicx}
\usepackage{booktabs}
\usepackage{amsfonts}
\usepackage{hyperref}
\usepackage{lmodern}
\usepackage{amssymb}

\newcommand{\nn}{~,\nonumber\\}
\newcommand{\bes}{\begin{equation*}}
\newcommand{\ees}{\end{equation*}}
\newcommand{\be}{\begin{equation}}
\newcommand{\ee}{\end{equation}}
\newcommand{\bea}{\begin{eqnarray}}
\newcommand{\eea}{\end{eqnarray}}

\begin{document}
\begin{titlepage}

\title{The MIXMAX random number generator}
\maketitle
\begin{center}
\author{Konstantin G. Savvidy\footnote{ksavvidis(AT)gmail.com}  \\Department of Physics \\and\\Center for Transcriptional Medicine,\\Nanjing University, Nanjing, China} 
% School of Medicine, 
\begin{abstract} 
In this note, we give a practical solution to the problem of determining the maximal period of matrix generators of pseudo-random numbers which are based on an integer-valued unimodular matrix of size NxN 
known as MIXMAX and arithmetic defined on a Galois field GF[p] with large prime modulus p. The existing theory of  Galois finite fields is adapted to the present case, and necessary and sufficient condition to attain the maximum period is formulated. Three efficient algorithms are presented. First,  allowing to compute the multiplication by the MIXMAX matrix with O(N) operations. Second,
to recursively compute the characteristic polynomial with O(N$^2$) operations, and third, to apply  skips of large number of steps S to the sequence in O(N$^2$ log(S)) operations.  It is demonstrated that the dynamical properties of this generator dramatically improve with the size of the matrix N, as compared to the classes of generators based on sparse matrices and/or sparse characteristic polynomials.
% to recursively compute the characteristic polynomial with O(N$^2$) operations, and third, to apply  skips of large number of steps S to the sequence in O(N$^2$ log(S)) operations.  It is demonstrated that the dynamical properties of this generator dramatically improve with the size of the matrix N, in contradistinction to the classes of generators based on sparse matrices and/or sparse characteristic polynomials.
%, in contradistinction to the classes of generators based on Fibonacci-like recurrences which have sparse matrices and/or sparse characteristic polynomials such as  those  in the GFSR family and especially those based  on primitive trinomials, which get worse with the increasing order. This property does not seem to be related to the smallness/largeness of the modulus.  
Finally, we present the implementation details of the generator and the results of rigorous statistical testing.
\end{abstract} 
\end{center}
%\date{\today}

\end{titlepage}

\section{Introduction}
In \cite{yer1986a} it was proposed that k-mixing systems of Kolmogorov  \cite{kolmo, rokhlin, franklin} may serve as a suitable random number generator. The particular system chosen was the one realizing linear automorphisms of the unit hypercube in $\mathbb{R}^N$:
\be
\label{eq:rec}
u_i(t+1) = \sum_{j=1}^N A_{ij} \, u_j(t) ~\textrm{mod}~ 1
\ee
where $u \in [0,1)$.
For the purposes of generating pseudo-random numbers with this method, one chooses the initial vector $u(0)$,  called the ``seed",  with at least one non-zero component.

The entries of the matrix are integers: $A_{ij} \in \mathbb{Z}$ and
subject to the following two conditions \cite{kolmo, rokhlin, franklin, yer1986a} on the defining matrix $A$ :
\begin{itemize}
\item $ \det A =1$
\item the eigenvalues $\lambda_k$ of A must not lie on the unit circle, $|\lambda_k| \neq 1$ for all $k=1...N$.
\end{itemize}

The first condition assures that the map defines a volume preserving automorphism.  When the automorphism is viewed as defining a dynamical system with discrete time $t \in \mathbb{Z}$, then the second condition assures that nearby trajectories diverge exponentially. % except on a stable manifold of zero measure. 
There exists an everywhere dense, but discrete set of periodic trajectories all of which are unstable, and whose number as a function of the period $\tau$ asymptotically goes like $ e^{h\tau}/\tau$, where $h$ is the Kolmogorov entropy of the system \cite{kolmo,anosov}:
\[
h = \sum_{k: |\lambda_k|>1} \log |\lambda_k|
\]
The auto-correlation decay time $\tau_0$ is related to the entropy as $\tau_0 \le 1/h$.
In this way, a connection is made between the chaotic dynamics of the system, its set of periodic trajectories and the entropy. In particular, the auto-correlation time is directly related to the empirical notion of randomness of the sequence. The second condition above states that none of the eigenvalues should lie precisely on the unit circle, and moreover the formula for the entropy demonstrates that it is desirable that most of the eigenvalues of the matrix should lie as far as possible away from it. 

A particular matrix chosen in \cite{yer1986b} was defined for all $N \geq 3$:
\be
\label{eq:matrix}
A = %\hbox{
   \begin{pmatrix} % or pmatrix or bmatrix or Bmatrix or ...
%2 & 3 & 4 & 5 & ... &&    N  &  1\\
%1 & 2 & 3 & 4 & ... &&   N-1 & 1\\
% &  & ...&&&&\\
%1 & 1 & 1 & 1 & ... & 2 & 3+s & 1  \\%  <- this is not a typo, the second entry from the right is 2, not 3!
%1 & 1 & 1 & 1 & ... & 1 & 2 & 1\\
%1 & 1 & 1 & 1 & ... & 1 & 1 & 1\\
      1 & 1 & 1 & 1 & ... &1& 1 \\
      1 & 2 & 1 & 1 & ... &1& 1 \\
      1 & 3+s & 2 & 1 & ... &1& 1 \\
      1 & 4 & 3 & 2 &   ... &1& 1 \\
      &&&...&&&\\
      1 & N & N-1 &  N-2 & ... & 3 & 2
   \end{pmatrix}
   %}
\ee
The MIXMAX matrix contains integer numbers and is defined recursively, since the matrix of size $N+1$ 
%shares everything except the last column and last row with the Matrix of size N-1. 
contains in it the matrix for size $N$.  The only variable entry in the matrix is $A_{32} =3+s$ where $s$ is some small ``magic" integer, in many cases $s=-1$ or  $s=0$. For those $N$ for which one or more eigenvalues lie on the unit circle for some $s$, one can choose another $s$.

\begin{figure}[htbp]
   \centering
   \includegraphics[width=14cm]{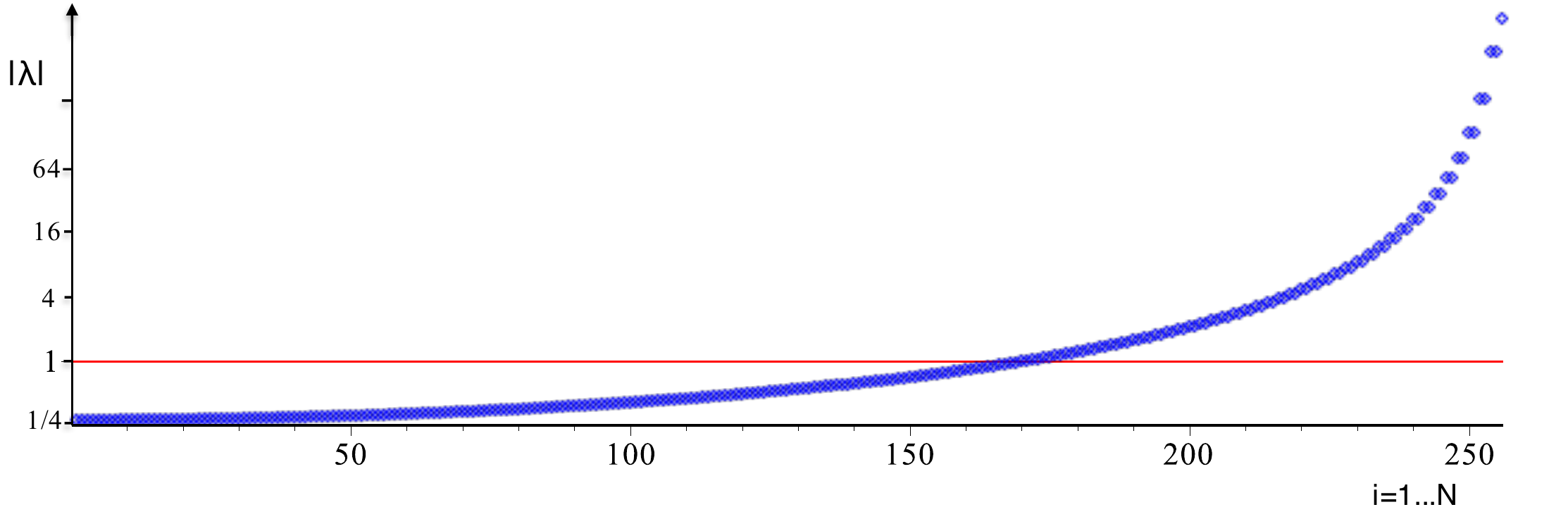} % requires the graphicx package
   \caption{The absolute value of the eigenvalues of the MIXMAX matrix for N=256 (logarithmic scale). Almost all of the  eigenvalues are far away from the unit circle.}
   \label{fig:mixmaxev}
\end{figure}

The eigenvalues of the MIXMAX matrix are widely dispersed for all $N$, see Figure  \ref{fig:mixmaxev}. Thus, the spectrum of this system is multi-scale, with trajectories exhibiting exponential instabilities on all time-scales \cite{yer1986a}. 

 Empirical evidence suggests that the largest eigenvalue appears to grow at least linearly with $N$, but we have been unable to obtain a strict bound on it. However, Kolmogorov's entropy can be more conveniently calculated using the small eigenvalues as follows. Since the product of all of the eigenvalues is equal to the determinant,
\[
\prod_{k} \lambda_k = \prod_{k} |\lambda_k| = 1 ~, \textrm{~~and~~~} \sum_{k} \log |\lambda_k| = 0 ~,
\]
the entropy can be calculated equally well  using the eigenvalues which are less than one by absolute value:
\[
h =  - \sum_{k: |\lambda_k|<1} \log |\lambda_k|
\]
i.e. the entropy is equal to the area under the upper branch of the curve in Fig. \ref{fig:mixmaxev} and also is equal to the area under the lower branch of the curve, taking into account that the vertical axis is already set to the logarithmic scale.

None of the eigenvalues of the Matrix A is smaller by absolute value than 1/4, regardless of N, and a large number of them tend to cluster just above 1/4. Therefore the Kolmogorov entropy can be strictly bound from above by bounding the area under the lower branch of the curve as follows: 
\[
h < - N \log |\lambda_{min}|  < N \, \log(4)~.
\]

We have also obtained an heuristic formula for the entropy which is  more precise. The actual number of eigenvalues lesser than one by absolute value is asymptotically $2/3 \, N$, and logarithm of their value is observed to lie close to a parabola:
\be \log(\lambda_k) \lesssim \log(4)\left(-1 + \left(\frac{3}{2N}\right)^2 \, k^2\right) ~~\textrm{for}~ k=1...\frac{2N}{3}.
\ee
Adding these up, we get  an approximate asymptotic formula, which also satisfies the strict bound above:
\[
N \log(4) > h \gtrsim 4/9\,N \, \log(4)
\]
This estimate appears to be reasonable, i.e. for $N=256$ the actual value is $h \simeq 164.4$ versus our estimate of $h \gtrsim 157.7$. 
% indicating that the corrections are positive and $O(\log(N))$.

\begin{figure}%[htbp]
   \centering
   \includegraphics[width=8cm]{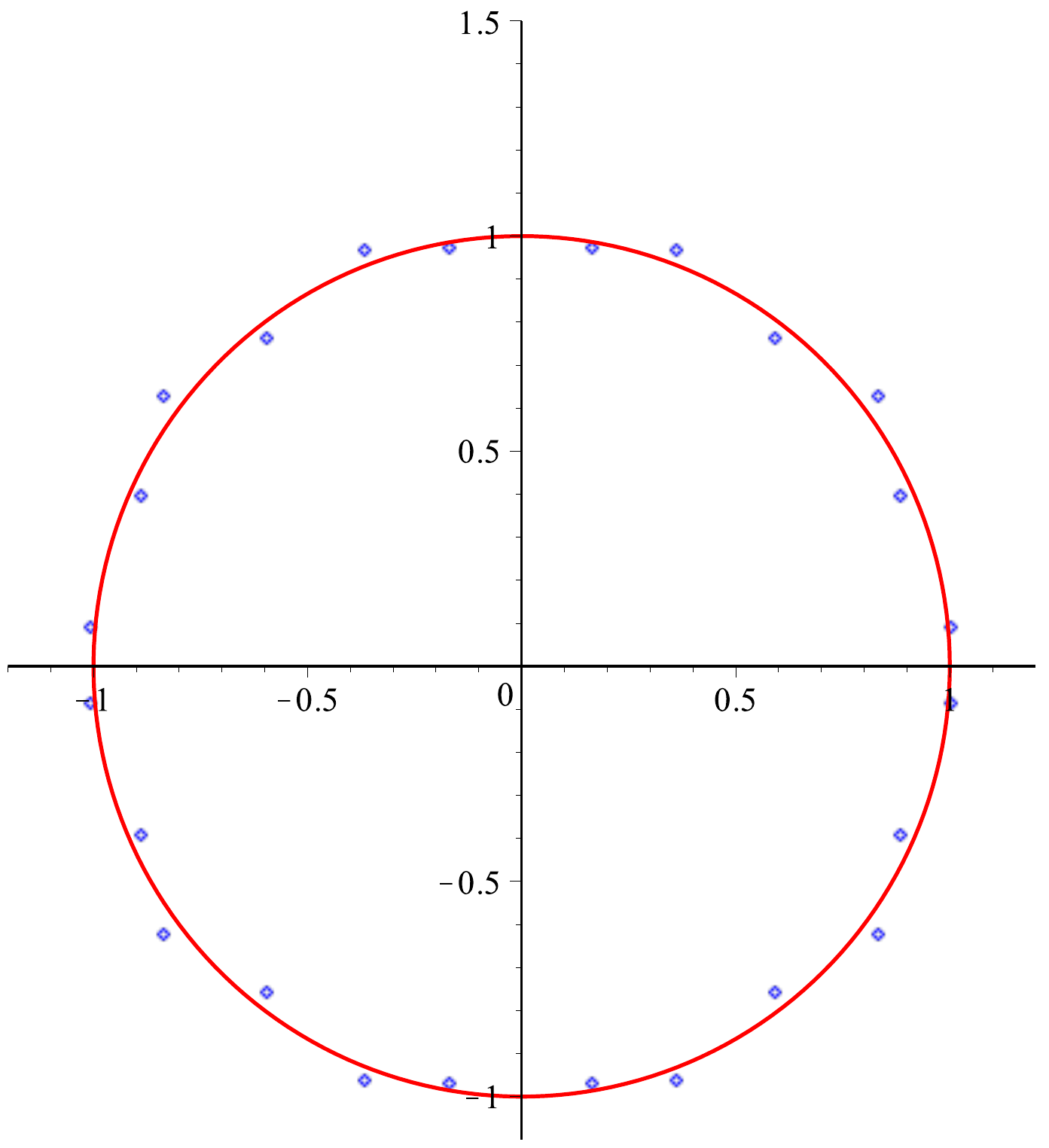} % requires the graphicx package
   \caption{Eigenvalues of the defining matrix of the RCARRY generator \cite{carry} all lie close to the unit circle. The eigenvalues closest to the circle have $|\lambda| \approx 1.0085$, the farthest $|\lambda| \approx 1.043$. In order to overcome the strong correlations exhibited by RCARRY as a result of this property, L\"uscher proposed to skip a predefined number of steps in the sequence \cite{ranlux}.}
   \label{fig:rcarryev}
\end{figure}

\begin{figure}%[htbp]
   \centering
   \includegraphics[width=14cm]{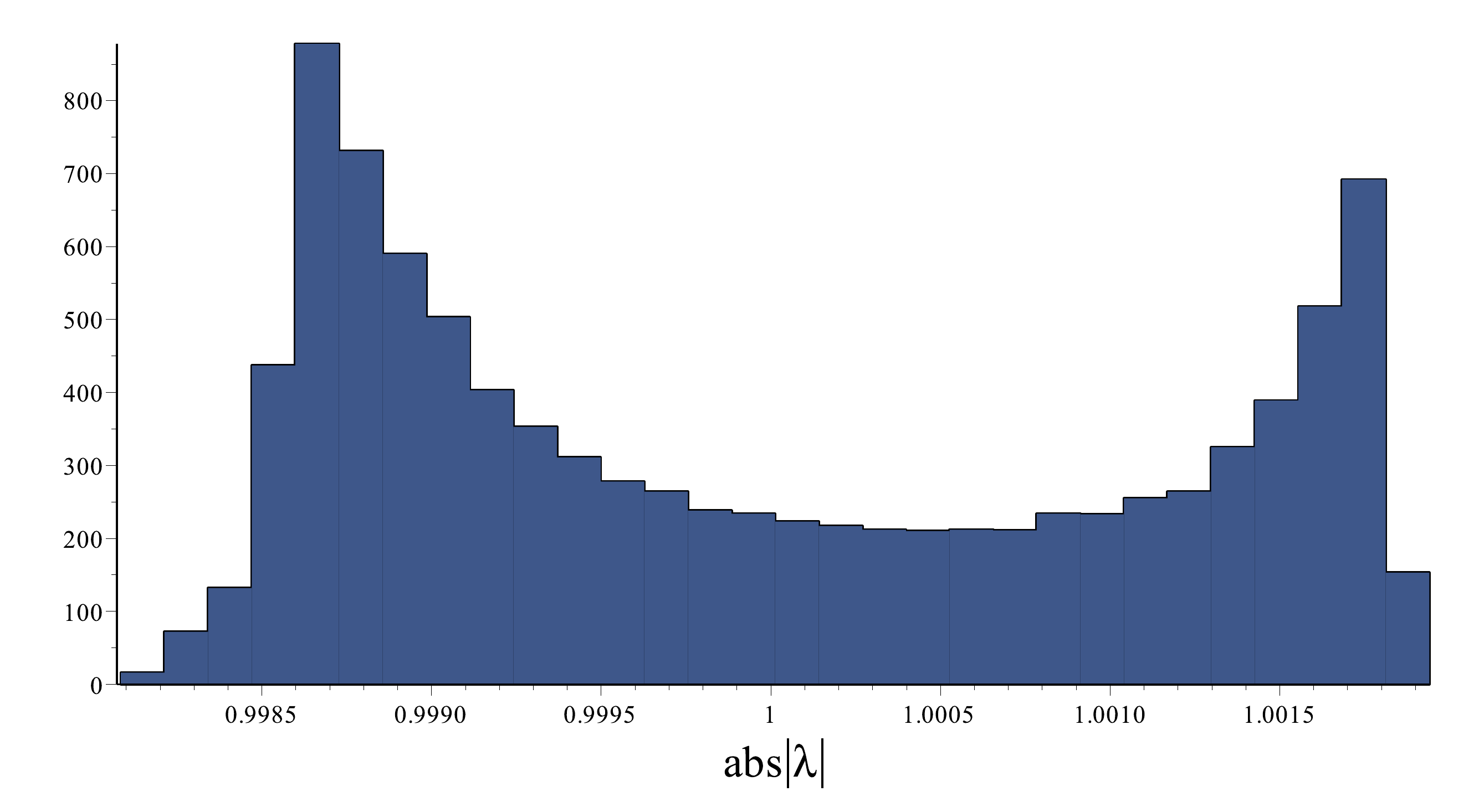} % requires the graphicx package
   \caption{Eigenvalues of the defining matrix of the Mersenne Twister generator \cite{carry} all lie close to the unit circle. The smallest (by absolute value) eigenvalue has $|\lambda| \approx 0.9982$, the largest $|\lambda| \approx 1.0019$. }
   \label{fig:mtev}
\end{figure}

Figure \ref{fig:rcarryev} demonstrates graphically that some of the other popular generators do not satisfy the second requirement for randomness. In the case of RCARRY \cite{carry} which is a slight modification of a Fibonacci-like recurrence modulo $2^{24}$, this point has been made before by L\"uscher \cite{ranlux}, and its failure was related to the weak mixing properties of its underlying matrix. Unfortunately, 
 the Mersenne Twister (MT) \cite{MT}  has not been studied  from
this point of view. Our preliminary investigation indicates that the real eigenvalues of its characteristic polynomial are distributed very close to the unit circle, with the largest being less than $|\lambda_{max}| < 1.0019$, see Fig. \ref{fig:mtev}.
% the real ev is -1.00177454624679475432923928322045118118903928291174955 
% Another root is at .9983908574 + 0.001763611039*I , abs(l) = 0.9983880465
% better .9983864 + 0.00177633632*I 
% j=33, at -.998770871869-0.01017660238012364*I  , abs=0.9988227159
% j=32, at -.987725115982-.1657118279*I
% so far largest found is .980307114729048887401503969727203153382932633872298718+
%                         .206949686822121067190767824150962837119806219959413781*I
% with abs= 1.0019132757200927558595573007845272507714580309046
In this sense, the underlying dynamical system of the Mersenne Twister has one order of magnitude less entropy than RCARRY.  
This is related to the singular flaw acknowledged by the authors of the Mersenne Twister, which is that MT has a very long recovery time when it is seeded by a vector containing mostly zeros: the output is observed to be non-random even after outputting a million values. In this instance, the divergence of some trajectory is observed away from the origin. In fact, from the dynamical system point of view \emph{this is not merely a manifestation of an unlucky initialization}, since any two nearby trajectories of the MT diverge very slowly and this is ultimately what causes the failure of MT in the statistical tests.

The total entropy of the MT system is approximately $h \approx 4.8$. On the basis of the known behavior of RCARRY and our investigation of the MIXMAX, it appears that an entropy of at least $h \gtrsim 50$ is required for a generator to have a sufficiently random trajectory. In light of this, it is perhaps not surprising that the Mersenne Twister fails many tests in its pure form, and still fails some tests even when some additional tempering of the output is applied.

One can even conjecture that the sequence produced by the Mersenne Twister would equally benefit from the method of skipping proposed by L\"uscher. However, the required amount of skipping would likely be prohibitively expensive. Moreover, it is clear that the recurrences based on primitive trinomials of even higher order such as those found in \cite{brent} would exhibit even longer correlations.

\section{Discrete case}
In a typical computer implementation, the recursion \eqref{eq:rec} can be used to generate uniform random variables on the unit interval directly in floating point hardware. However, since actual floating point hardware has finite precision the sequence will tend to lie on a rational sublattice, with components of the vector $x_i$ being multiples of $2^{-b}$ where $b$ is the number of bits available for the mantissa of the floating point unit, typically $b=53$ on current computers. There are at least two drawbacks to this scheme. First, the period of the generator so realized will strongly depend on the initial seed. Second, the operation of truncation of the integer part which is indicated in eq. \eqref{eq:rec} may not be particularly efficient. 

We now consider this from another point of view \cite{mixmaxGalois}. If the initial vector happens to have rational components $u_i=m_i/n_i$, where $m$ and $n$ are natural numbers, then all subsequent vectors will remain on the rational sub-lattice $u_i=a_i/p$ where $p$ is the least common multiple of the $n_i$. In this case it is convenient to represent $u_i$ by its numerator $a_i$ in computer memory, or even better, to simply define the same recursion in terms of $a_i$:
\be
\label{eq:recP}
a^\prime_i = \sum_{j=1}^N A_{ij} \, a_j ~\textrm{mod}~ p
\ee
In the rest of the paper we shall not indicate the modular operations explicitly, and use the equal sign in the sense of equivalence modulo $p$.

Random number generators of this general form, without reference to the underlying automorphism of the torus and the dynamical systems theory were proposed by Tahmi \cite{tahmi} and Niki \cite{Niki} and were extensively studied by Grothe, Niederreiter and others \cite{grothe,mixmaxGalois,nied,lnbook}. These authors established the connection to the theory of finite extended Galois fields, and obtained results regarding the period of the recursion when the modulus $p$ is a prime.

%Therefore it is natural to consider $p$ which is a prime number from the beginning, f
On a given computer architecture, $p$ could be chosen as the largest ordinary or Mersenne prime lesser than the maximum unsigned integer.  A distinct class of recursions results if $p=2$, in which case the GF[2] arithmetic may be naturally realized on the available computer architectures via the bitwise XOR operation.

We now summarise the known mathematical facts about such linear matrix recursions, modulo a prime number $p$. The sequence of vectors $a(t)$ generated by the recursion is necessarily periodic, and the period cannot exceed the cardinality of the nonzero elements of the finite vector field, which is equal to $p^N-1$. If, and only if, the characteristic polynomial $P(x)$ of the matrix $A$, $P(x) = (-1)^{N} \det |A - x \, \mathbb{I} |$ is primitive in the extended Galois finite field $GF[p^N]$, then the period of the recursion attains its maximal value, $p^N-1$, independent of the seed. The necessary and sufficient conditions for this are well known \cite{lnbook}. One of the necessary conditions is the following:
\begin{itemize}
\item  the free term of the polynomial $p_0 = P(x)\vert_{x=0}$, also equal to the determinant of the matrix  $p_0 = \det A $, is a primitive element of the Galois field GF[p]
%\item or else, $p_0=1$ and $p=2$
\end{itemize}
The determinant of the MIXMAX matrix is equal to one, since it is one of the conditions for Kolmogorov k-mixing to occur, and for the mapping to be an automorphism. Therefore, as it was noted in \cite{mixmaxGalois}, the recursion defined by \eqref{eq:recP} cannot attain the maximal period (unless $p=2$). In the next section we investigate the maximal possible period of the sequence \eqref{eq:recP} for $p \neq 2$, and generalize the notion of primitive polynomials to the required case.
 
\section{Maximal period sequence}
%It was found at first empirically, that the maximum possible period of the sequence defined by \ref{eq:recP} is $q=\frac{p^N-1} {p-1}$.
\renewcommand{\mod}[1]{}

A key insight which allows to determine the maximal period of the sequence defined by \eqref{eq:recP} is that, in the case that the characteristic polynomial primitive,
\[ A^q = p_0~ \mathbb{I}~~\textrm{ for}~~  q=\frac{p^N-1} {p-1} ~,\]
where as before  $p_0$ is the free term of the characteristic polynomial of A and is equal to the determinant of A. Therefore, the powers of the matrix A which are multiples of q are diagonal:
\[ A^{mq} \mod p = \left(A^{q}\right)^m \mod p = p_0^m ~\mathbb{I} \]
Since the MIXMAX matrix has $p_0=1$, therefore the maximum possible period is equal to $q$.
 %The necessary and sufficient conditions for this to happen were found to be:
We define two necessary and sufficient conditions for the period $\tau$ of the unimodular matrix recursion  to attain its maximum possible value and be equal to $q$:
\begin{enumerate}
\item[\bf{1.}] $A^q \mod p = \mathbb{I}$
\item[\bf{2.}] $A^{q/r} \mod p \neq \mathbb{I}$ for any r which is a prime divisor of q % , $r=q/q_i \neq 1 $
\end{enumerate}
%\end{itemize}

In practice, it is convenient to calculate the modular exponentiation of a matrix using the theorem of Hamilton and Cayley. Since $P(A)=0$, we can always reduce some high power of the matrix in terms of powers up to $N-1$:
\be
A^m \mod p = e_{N-1} A^{N-1} + ... + e_2 A^2 + e_1 A + e_0~,
\label{eq:hamcayley}
\ee
where the coefficients can also be obtained by polynomial algebra:
\be
E(x) = x^m ~ \textrm{mod} ~ P(x) \mod p = e_{N-1} x^{N-1} + ... + e_2 x^2 +e_1 x + e_0~.
\label{eq:polmod}
\ee

%Therefore, the first condition above, and therefore also the irreducibility of the characteristic polynomial is checked simply by making sure that $ x^q \mod P(x) \mod p = p_0$. 

For the first condition to be satisfied, it is necessary and sufficient that the characteristic polynomial is irreducible: $  P(x) ~\textrm{mod} \, Q(x) \mod p \neq 0$ for any polynomial $Q(x) \neq 0$. The second condition can be checked directly, by computing $x^{q/r} ~\textrm{mod} \, P(x) \mod p$ for all $r$ which are prime divisors of $q$. If it is violated for some $r$, then the actual period of the sequence is equal to the q divided by the least common multiple of any such r. If both conditions are satisfied, then we may call the corresponding characteristic polynomials ``quasi-primitive".
%The second condition can be checked directly, which is feasible in practice if an integer factorization of $q$ is available. 
As it should be obvious from the definition, the only condition which is relaxed compared to the ordinary, primitive polynomials, is that the free term of the characteristic polynomial $p_0$ is not required to be the primitive element of the base field GF[p]. In the particular case of interest to us, the matrix A in \eqref{eq:matrix} has $p_0=\det A = 1$.
% so it follows that a unimodular matrix cannot have a primitive characteristic polynomial unless $p=2$.
% This in turn results in $A^q \mod p = I$ and therefore also  $A^{qm} \mod p = I$ for all natural $m$.

From here, it follows that the period of the sequence is equal to $q$, and is independent of the seed. Moreover, there are precisely $p-1$ disjoint sequences which together fill up the entire space of states: $q \,(p-1) = p^N-1$. 
%For practical purposes, this may allow to easily obtain parallel streams of random numbers by allowing each instance of the generator to sit on a different, non-overlapping sequence. 
It appears to be a difficult mathematical problem to decide whether two given vectors belong to the same or different sequence. However, one may jump from one sequence to another non-overlapping sequence by multiplying the initial vector $a(0)$ or indeed any subsequent vector $a(t)$ by a number $c$, subject to some conditions on $c$.
% so long as $c$ is smaller than  the order of any root of unity in $GF(p)$ of an order which divides $q$. 
We will be able to lift this minor restriction, and make a more complete description of the entire set of disjoint trajectories in the next section.

%This annoying situation with roots of unity can be avoided as follows. It appears that the necessary and sufficient additional requirement for 
% \[ A^i \neq c  \, \mathbb{I}, ~\textrm{for any } ~ c \in GF[p] ~~\textrm{for all} ~~ i<q \]
%  is to demand that $q$ and  $p-1$ are relatively prime, $\gcd(q, p-1) = 1$. 
%If this takes place, and all the previously imposed conditions of the  are also satisfied, then all of the $p-1$ disjoint trajectories are related to each other by multiplication by a number. 

All of the above preceding mathematical statements can be justified on the basis of the theory of finite fields, specifically most of the necessary proofs can be obtained by applying the method developed in the Lemma 3.17 in the book by Lidl and Niederreiter \cite{lnbook}. % p 117 in russian edition 

Furthermore, it is possible to carry out the check in the second condition only if the full integer factorization of $q$ is available. In the contrary case, we may suppose that $q$  is only partially factorized:
\be
q = r_1^{\alpha_1} \, ... \, r_k^{\alpha_k} ~~ c_1 \, ... \, c_m~~,
\ee
where $r_i$ are the prime divisors, $\alpha$ their multiplicities and $c_i$ are the composite co-factors of $q$. Further, suppose that it is known that all of $c_i$ have no divisors smaller than some common lower bound $u$. Assuming that the second condition holds for all of $r_i$ and $c_i$, we can obtain the strict bound on the period of the sequence:
\[ q \ge \tau >  r_1^{\alpha_1} \, ... \,r_k^{\alpha_k} \, u^m \]

If $u$, the lower bound on the unknown divisors is sufficiently high, then one can state with very high certainty that nevertheless the period $\tau$ is in fact equal to $q$, with probability that goes asymptotically as $1- u^{-1} \to 1$.

As a practical matter, it is contemplated to use $p=2^{61}-1$, the largest Mersenne number that fits into an unsigned integer on current 64-bit computer architectures. Fortunately, this is enough to produce double-precision floating point values which are random down to their lowest bit. 

The complete integer factorization of $q = \frac{p^N-1} {p-1}$ for some N is available. Typically, it is easier to factorize for some N, and then reuse the same factors for $N=kN$ for some small $k$. The greatest benefit from these algebraic factorizations results if $N$ is a product of some small primes, for example a primorial. %The largest currently factorized number $q$ is that for $N=120$. 
It is not difficult to find all of the small divisors for any N by means of the Elliptic Curve Method (ECM), this is useful for reasons explained above.

We have made searches for irreducible polynomials for various values of $N$ and the ``magic" number $s$, which are summarized in Table \ref{tbl:Ns}. The period of the sequence for all of the given parameters is astronomical, and cannot be exhausted even if the number of parallel instances of the generator is itself astronomical. Nevertheless, the generator fails statistical tests for small N. Therefore, the period by itself cannot be considered a measure of quality of a generator. On the other hand, the entropy and the empirical randomness of the generator gets uniformly better with N. For all $N > 64$ and $h \gtrsim 50$ which we have tested, the generator passes all tests. 

   We recommend to choose the value of N based on the problem at hand. From the dynamical systems point of view, Monte Carlo simulation of a Markov Chain (MCMC) should be done with a random number generator whose auto-correlation time $\tau_0 = 1/h$ is much smaller than the auto-correlation time of the Markov chain. This is easily satisfied by choosing a sufficiently large size N of the generator matrix. On the other hand, difficulties also arise when simulating a system with very long auto-correlation time, for example the Ising model with the Metropolis method near the critical point. In this case, and other cases where the effective dimension of the Monte-Carlo integration is large, one should choose N which is larger than this effective dimension. Usually, this requirement is stated in terms of the equidistribution of the sequence. No generator can guarantee equidistribution in a dimension larger than the dimension of its internal state \cite{mars}.
   %, but one can try to achieve a more or else acceptable lattice structure.

% Requires the booktabs if the memoir class is not being used
\begin{table}[htbp]
   \centering
   %\topcaption{Table captions are better up top} % requires the topcapt package
   \begin{tabular}{@{} rcrlrcl @{}} % Column formatting, @{} suppresses leading/trailing space
      \toprule
      Size & Magic & Entropy & Period &                                         &    ~~q is & \\
      N    & $s$ & (lower bound)  & $\tau/q$   & $\approx \log_{10} (q)$ & fully factored & BigCrush \\ % Crush
      \midrule
      10   & $-1$  &    6.2       & 1/4       & ${165}$               & Yes & 33       \\ % (24)  --- s=-1 is q/4, s=-3 is Full
      16   & $6$    &   9.9       & 1/32      & ${275}$               & Yes & $>13$ \\ % (13)
      40   & $1$    &   24.6     & 1/4       & ${716}$               & Yes & 3          \\  % N=30 Passes, on BigCrush N=30 has (9)
      44   & $0$    &  27.1        & 1/4       & ${789}$               & No & 4            \\   
      60
              & $4$  & 37.0         & 1          & ${1083}$             & Yes & 2            \\  % (1 suspect on Crush)
      64   & $6$     & 39.4         & 1/8       & ${1156}$               & No & 1 (?)           \\
      88   & $1$     & 54.2         & 1/2       & ${1597}$               & No & Pass       \\
      256   & $-1$   & 157.7        & 1         & ${4682}$             & No & Pass            \\
      508   & $5$    &  313.0       & 1       & ${9309}$             & No & Pass             \\
      720   & $ 1 $  & 443.6        & 1       & ${13202}$          & No & Pass             \\
      1000 & $0$     & 616.1     & 1/20    & ${18344}$          & No &    Pass         \\
      1260 & $15$    & 776.3     & 1/2     & ${23118}$          & No &     Pass        \\
      3150 & $-11$   & 1940.8    & 1/12   & ${57824}$          & No &    Pass         \\
      \bottomrule
   \end{tabular}
   \caption{Table of properties of generators for different matrix size $N$ and special magic value $s$. For each N that we investigated, the period $\tau$ is given as a fraction of $q=(p^N-1)/(p-1)$. 
%For $N<1000$ we kept only those $N$ for which the period is equal to the maximal possible value, $q$. 
For cases where the full integer factorization of $q$ is known, unconditional guarantee can be given about the period of the sequence. In all cases the characteristic polynomial was proved to be irreducible by Pari/GP \cite{parigp}. The last column indicates whether the generator for that $N$ and special value $s$ passes the BigCrush suite of tests, and if not how many tests are failed. The case of $N=60$ uses a doubly special matrix which has two entries modified: $a_{32}=a_{54}=3+s$. It is seen that the generator gets uniformly better with N until it passes all tests. The most discriminative test for this family of generators appears to be the classic Gap test. On this test alone, the improvement with N is also evident, with progressively better p-values as N is increased, e.g. for N=64 the value of $\chi^2 \approx 372$ for 232 degrees of freedom with $\chi^2/dof \approx 1.6$ 
indicates only a marginal failure. For all $N > 64$ which we have tested, the generator passes all tests.}
   \label{tbl:Ns}
\end{table}

For $p=2$, $q$ may happen to be a prime, called Mersenne prime. In the past, useable random number generators could be constructed on the basis of some   of the  available Mersenne primes, such as $N=19937$. In many of these cases, a search is made for a primitive trinomial in GF[2] of order N, and a corresponding Fibonacci recursion is used to generate random bits (the GFSR family). Otherwise, as in the case of the Mersenne Twister, a sparse matrix is constructed of the almost-banded form whose characteristic polynomial is proved to be irreducible for some values of the magic entries. Since $q$ is prime, the second condition is satisfied automatically, and therefore the characteristic polynomial is truly primitive in these cases. 
% Unfortunately, the larger the N, the closer the roots of the characteristic polynomial get to 1 by absolute value. In fact, \emph{all} of the roots tend to lie very close to the unit circle in the complex plane for these generators. %For example, in the case of the Mersenne Twister ...

%The large modulus generators do not suffer from this problem. 
%Therefore it is very useful to be able to...

\section{The case of the prime period}
The contents of the previous section can be given an elegant extension  by the following observation:

\emph{if N is prime, then q may happen to be a prime number as well and therefore the period $\tau$ is guaranteed to be equal to $q$,   $\tau=q$, if and only if the characteristic polynomial is irreducible}. 

%Let us consider  elements of the extended Galois field $GF(p^N)$ to be identical if they are multiples of each other with respect to the base field GF(p):
%\be
%\alpha_1 \equiv \alpha_2 ~~\textrm{iff} ~~\exists \, c \in GF(p), ~\textrm{such that} ~ \alpha_1 = c \alpha_2 %\mod p
%\label{eq:id}
%\ee
%The equivalence classes so defined are elements of a new finite field, which we may call the quotient of $GF(p^N)$/GF(p), or the affine extension of the base Galois field. This definition of an affine field is equally applicable to the case of the affine vector field of dimension N and the affine polynomial field of polynomials of order N, both subject to the identification of non-zero elements according to \eqref{eq:id}.
%
%We may choose to call this new field in both cases as the \emph{Galois Affine Extended Field} (GAffE Field).  This finite field has precisely $q = \frac{p^N-1}{p-1}$
%non-zero elements, since each non-zero element lies in the equivalence class with precisely $p-1$ elements (including itself). The zero element lies in its own equivalence class by itself.
%
%It is obvious that the matrix A implements the maximal period recursion in the affine vector field if and only if its characteristic polynomial is a primitive element of the affine polynomial field. 

%Unfortunately, the reverse is not true, namely not every quasi-primitive polynomial is a primitive element of the affine field.

There are no algebraic factors of $q$ if $N$ is an odd prime. %There may exists additional nontrivial factors of $q$, called Arifeuillian factors, when $N = p$ ??? .
There are no trivial factors if $N$ is co-prime with $p-1$. 
Other than this, naive considerations indicate that the \emph{a priori} probability that some particular $q$ is a prime is finite, and goes like $\frac{1}{N \log(p)}$. For $p=2$,  if $2^N-1$ happens to be a prime it is called a Mersenne prime. Primes of the form $\frac{p^N-1}{p-1}$ for some prime $N$ and a small prime base $p$  are called {``repunit" primes}, and it is a natural extension of the notion of Mersenne primes to $p \neq 2$. Our $q$  are  ``repunit" numbers in base $p$, since $q= 1 + p + p^2 + ... + p^{N-1}$ (a repunit number has all digits equal to one in some base, here $p$). However, the bases we use are very large, so it is somewhat unnatural to call it ``repunit", a more appropriate name might be ``affine-Mersenne" prime.
%Most importantly, we preserve the highly desirable property that if the cardinality of the non-zero elements of the  extended finite field is a prime, then irreducibility of the defining polynomial together with primitivity of its free term implies primitivity.

 The period $q$ may happen to be a prime also when $p$ itself is an ordinary Mersenne prime, but unfortunately, for $p=2^{61}-1$ we did not find any such N for which $q$ is a prime number. Not to be discouraged, we have made a non-exhaustive search for some other primes $p$, chosen for convenience to be just short of $2^{62}$ or $2^{63}$. The search has yielded the combinations of $p$, $N$ and $s$ for which $N$ and $q$ are prime, and the characteristic polynomial of the MIXMAX matrix (for such $s$) is irreducible. The second condition of the previous section is then trivially fullfilled and therefore the characteristic polynomial is quasi-primitive. The results are presented in Table \ref{tbl:prime}.

% Requires the booktabs if the memoir class is not being used
\begin{table}[htbp]
   \centering
   %\topcaption{Table captions are better up top} % requires the topcapt package
\small{
   \begin{tabular}{@{} rclrcl @{}} % Column formatting, @{} suppresses leading/trailing space
      \toprule
      Modulus & Size & Magic & Period            &     & \\
     p             &  N    & $s$      &    $\approx \log_{10} (q)$ &  BigCrush \\
%      \midrule
%$p<2^{63}$&&&&\\
%      \midrule
%%  9223372036854769163  &   17 & 0 &    303 & \\
%%  9223372036854775643  &   19 & 3 &    341 & \\
%%  9223372036854763057  &   31 & 2 &    568 & \\
%%  9223372036854762463  &   47 & 5 &    872 & \\
%%  9223372036854766261  &   53 & -3 &    986 & \\
%%  9223372036854736009  &   59 & -4 &   1099 & \\
%  9223372036854734281  &   61 & -1 &   1137 & $\geq1$ \\
%%  9223372036854340537  &   67 & -1 &   1251 & \\
%  9223372036854205259  &   67 & 0 &   1251 & $\geq1$\\
%  9223372036854753853  &   71 & 3 &   1327 & \\
%  9223372036854759529  &   97 & -7 &   1820 & \\
%  9223372036854758317  &  103 & -4 &   1934 & \\
%  9223372036854609833  &  127 & -1 &   2389 & \\
%  9223372036854176521  &  139 & 0 &   2617 & \\
%%  9223372036853751941  &  139 & 0 &   2617 & \\
%  9223372036854767881  &  211 &  27 &   3982 & \\
%  9223372036854682991  &  257 &  5 &   4855 & \\
%      \midrule
%$p<2^{62}$&&&&\\
      \midrule
  4611686018427341489  &   17 & 0 &    298 & 14 \\
  4611686018427246217  &   19 &  0 &    335 & 7 \\
  4611686018427365419  &   23 &  0 &    410 & 6 \\
  4611686018427297023  &   31 &  0 &    559 & 3 \\
  4611686018427370139  &   37 &  -1 &    671 & 2 \\
  4611686018427317411  &   43 &  -1 &    783 & 1 \\
  % 4611686018427211453  &   43 &  0 &    783 & ? \\
  4611686018427335557  &   47 &  -1 &    858 & 1 \\
  4611686018427262387  &   53 &  1 &    970 & 1 \\
  4611686018427084347  &   59 &  3 &   1082 & 1 \\
  4611686018426896543  &   61 &  0 &   1119 & 1 \\
% 4611686018427138193  &   67 &  3 &   1231 & ? \\
  4611686018427033523  &   67 &  2 &   1231 & 1 \\
  4611686018427208187  &   71 &  4 &   1306 & 1 \\
  % 4611686018426728387  &  127 &  -1 &   2351 & ? \\
  4611686018426592983  &  127 &  0 &   2351 & Pass \\
%  4611686018427372029  &  257 &  10 &   4777 & ? \\   % suspicious entry, reconfirm!!! %  4611686018427322771  &  257 &  10 &   4777 & ? \\
%  4611686018427190237  &  269 &  25 &   5001 & ? \\
9223372036853751941  &  139 & 0 &   2617 & Pass \\
  9223372036854661783  &  257 &  -4 &   4855 & Pass \\
%  9223372036854775417  &  359 &   7 &   6789 & \\

 \bottomrule
   \end{tabular}
   }
   \caption{Table of properties of generators for different matrix size $N$ and special magic value $s$. For all of these generators, the period $\tau$ is equal to $q$, $\tau=q=(p^N-1)/(p-1)$. 
%For $N<1000$ we kept only those $N$ for which the period is equal to the maximal possible value, $q$. 
Since $q$ is prime for all combinations of $p$ and $N$ in this Table, unconditional guarantee can be given about the period of the sequence. In all cases the characteristic polynomial was proved to be irreducible by Pari/GP \cite{parigp}. The last column indicates whether the generator for that $N$ and special value $s$ passes the BigCrush suite of tests, and if not how many tests are failed. It is seen that the generator gets uniformly better with N until it passes all tests. 
%For all $N > 64$ which we have tested, the generator passes all tests.
}
   \label{tbl:prime}
\end{table}

    In all of the cases  in Table \ref{tbl:prime} we have $\gcd(q, p-1) = 1$. Compared to most of the cases considered in the previous section, in Table \ref{tbl:Ns}, this gives the additional benefit: the matrix $A^m$ is not diagonal for any power of $m$ less than $q$:
\be
A^m \neq c\, \mathbb{I} ~~ \textrm{for any} ~~c=1...(p-1) ~ \textrm{and}~ m<q % \forall 
\ee
unlike many of those in Table \ref{tbl:Ns}. For example, whenever the  period $\tau$ is even,  we have
\be
A^{\tau/2} = - \mathbb{I} = (p-1) \, \mathbb{I}
\ee
or, more generally
\be
A^{\tau/r} = \sqrt[r]{1} \,  \mathbb{I}
\ee
for any $r$ which is a common divisor of $q$ and $p-1$ (since  $\sqrt[r]{1} \in GF(p)$ whenever $r || (p-1)$). Obviously, this annoyance does not occur when q is a prime number.

In practice, this means that one can jump from any one disjoint trajectory of the generator to another by multiplying the state vector by a number in the range $2...(p-1)$.

\section{Efficient Implementation}
In this section, we give for completeness the formulae which allow the efficient implementation of the generator in actual computer hardware.

First, we present the formula which allows the efficient calculation of the recursion. Given the vector $a$ with components $a_i$, $i=1...N$, a vector of partial sums $b$ is formed according to 
\bea
b_1 &=& 0 \nn
b_i &=& b_{i-1} + a_i ~~~ \textrm{for}~i=2...N~.
\label{rec1}
\eea
Then, the new vector is calculated:
\bea
 a^\prime_1 &\leftarrow& a_1 + b_N \nn
 a^\prime_i &\leftarrow& a_{i-1} + b_i~~ ~ \textrm{for}~i=2...N ~.
 \label{rec2}
 \eea
 Finally, the correction due to the magic value is applied
 \[  a^\prime_3 \leftarrow a^\prime_3 + s \, a_2 \]
It is obvious that the above is a unimodular linear transformation, but it is somewhat less trivial to see that \eqref{rec1} and \eqref{rec2} indeed implement the multiplication by the MIXMAX matrix of \eqref{eq:matrix}.

Next, we  present without proof the recursive formulae which allow the efficient calculation of the characteristic polynomial $P_N(x) $ of the MIXMAX matrix:
\bea
M_0 &=& 1 \nn
M_1 &=& 2\, x \nn
M_2 &=& 3\, x^2+x\nn
M_j &=& (2\, x)  \, M_{j-1}+ (1-x)\, x\, M_{j-2}
\eea
and finally
\be
P_N(x) = -x \, \Big[ (2\, x+ s)\, M_{N-3} + (1-x)\, (x+ s)\, M_{N-4} \Big]  + (x-1)^N
\ee

Finally, the formulae \eqref{eq:polmod} and  \eqref{eq:hamcayley} allow the realization of skipping. Fast application of the skipping matrix  $A^m$ to the state vector $a(t)$ is possible without actually computing or storing it. Instead, one pre-computes and stores the coefficients $e_i,~ i=0...(N-1)$ of the polynomial $E(x)$ defined in 
\eqref{eq:polmod}, for example for a modest set of $m$ which are powers of two up to 1024: $m=2^j, ~~j=0...1024$.  %In what follows, all equalities and assignments are understood to be in the Galois field, with modulus $p$.

According to \eqref{eq:hamcayley}, 
\be
a(t + m) = A^m \, . \, a(t) = E(A) \, . \, a(t) = \sum_{i=0}^{N-1} e_i \, A^i \,. \, a(t) = \sum_{i=0}^{N-1} e_i \,  a(t+i)
\ee
Furthermore, skipping is additive since the different powers of the matrix A commute. Therefore, it is possible to apply skipping by an arbitrary integer $S<2^{1024}$  by decomposing it as a sum of powers of two, $S= \sum 2^j b_j$, where $b_j$ is the $j$th bit of $S$. Then one simply scans over $j$, and applies the above formula recursively for all nonzero bits of $S$ with $m=2^j$.

\section{Conclusion and Acknowledgements}

The MIXMAX random number generator is currently made available by the author in  a portable implementation  in the C language at hepforge.org \cite{hepforge}. We urge the reader to consult the README file included in the distribution at hepforge.org for details on the use of the generator. Also, an experimental implementation intended as part of a development version of the C++ ROOT library from CERN exists at \cite{root}.

 The generator outputs in double precision natively, with all 53 bits random. The speed of the generator compares favorably to other generators currently available: in our tests it is significantly faster than RANLUX in 24-bit and 48-bit precision and still somewhat faster than the 32-bit precision Mersenne Twister. Also, among the generators with a large state space it is, to our knowledge, unique in offering efficient skipping. Large skipping capability allows to seed the generator for a large Monte-Carlo simulation on multiple clusters/CPUs with the absolute mathematical guarantee that the generated streams do not collide or overlap.
 
I would like to thank  Jan Ambjorn, Fred James and George Savvidy, the three people whose encouragement is chiefly responsible for my persistence over the years with implementing the generator in software and studying its theoretical properties.  I thank J. Apostolakis, J. Harvey and L. Moneta at CERN for useful discussions.

\vfill

\end{document}